\documentclass{article}
\usepackage[preprint]{spconf}
\usepackage{amsmath,graphicx}

\usepackage{cite} 
\usepackage{xcolor}
\usepackage{amsmath,amssymb,amsfonts}
\usepackage{booktabs}
\usepackage{array}
\newcolumntype{M}[1]{>{\centering\arraybackslash}m{#1}}

\copyrightnotice{\copyright\ IEEE 2022}
\toappear{To appear in {\it Proc.\ ICASSP 2022}}

\title{MANNER: Multi-view Attention Network for Noise Erasure}

\name{Hyun Joon Park, Byung Ha Kang, Wooseok Shin, Jin Sob Kim, Sung Won Han* \thanks{This research was supported by Brain Korea 21 FOUR. This research was also supported by Korea University Grant (K2107521) and a Korea TechnoComplex Foundation Grant (R2112651).}}
\address{School of Industrial and Management Engineering, Korea University, Seoul, Republic of Korea}

\begin{document}
\maketitle

\begin{abstract}
In the field of speech enhancement, time domain methods have difficulties in achieving both high performance and efficiency. Recently, dual-path models have been adopted to represent long sequential features, but they still have limited representations and poor memory efficiency. In this study, we propose Multi-view Attention Network for Noise ERasure (MANNER) consisting of a convolutional encoder-decoder with a multi-view attention block, applied to the time-domain signals. MANNER efficiently extracts three different representations from noisy speech and estimates high-quality clean speech. We evaluated MANNER on the VoiceBank-DEMAND dataset in terms of five objective speech quality metrics. Experimental results show that MANNER achieves state-of-the-art performance while efficiently processing noisy speech.
\end{abstract}

\begin{keywords}
multi-view attention, speech enhancement, time domain, u-net
\end{keywords}

\section{Introduction}
\label{sec:intro}

Speech enhancement (SE), which is the task of improving the quality and intelligibility of a noisy speech signal, has been widely used in many applications, such as automatic speech recognition and hearing aids. Recently, researchers have studied deep neural network (DNN) models for SE, as DNN models have shown powerful noise reduction ability in complex noise environments compared to statistical methods.

DNN models in SE are divided into time and time-frequency (T-F) domain methods. T-F domain methods \cite{choi2019phaseaware, fu2019metricgan, fu2021metricgan+, yin2020phasen, zheng2020interactive} estimate clean speech from the spectrogram created by applying the short-time Fourier transform (STFT) to a raw signal. Although the spectrogram contains the time and frequency of the signal, some limitations have been pointed out to use it \cite{wang2021tstnn, rethage2018wavenet}. T-F domain methods need to address both magnitude and phase information, thus increasing the model complexity. In addition, it is challenging to handle complex values for estimating complex-valued masks.

Recently, researchers have studied time domain methods \cite{pandey2020dual, defossez2020real, wang2021tstnn, macartney2018improved, rethage2018wavenet, pascual2017segan, pandey2020densely, pandey2021dense, hsieh2020wavecrn}, which directly estimate clean speech from the raw signal because the raw signal implicitly contains all of the signal’s information. Among them, \cite{defossez2020real, macartney2018improved, pandey2020densely, pandey2021dense} adopted a U-net \cite{ronneberger2015u} based architecture, which is utilized for efficient feature compression. However, it is not effective for representing the long sequence of the signal owing to its limited receptive field.

In contrast, dual-path models were adopted by \cite{luo2020dual, chen2020dual, subakan2021attention} to represent the long sequence of the signal in speech separation. They considered the long sequential features by dividing the signal into small chunks and repeatedly processing local and global information. In SE, \cite{wang2021tstnn, pandey2020dual} also applied dual-path models, but they are not efficient in terms of memory usage because they maintain the long signal length during training. In addition, the repeated feature extraction by dual-path processing on a small channel size results in limited representation and lower performance.

In this study, we propose an efficient speech enhancement model, Multi-view Attention Network for Noise ERasure (MANNER), in the time domain. MANNER, based on U-net, compresses the enriched channel representations with convolution blocks. The multi-view attention block enables the estimation of clean speech by emphasizing the channel and long sequential features from each view. A comparison of results on the VoiceBank-Demand dataset suggests that MANNER achieves state-of-the-art performance with high inference speed and efficient memory usage.

\begin{figure*}[htp]
\centerline{\includegraphics[scale = 0.5]{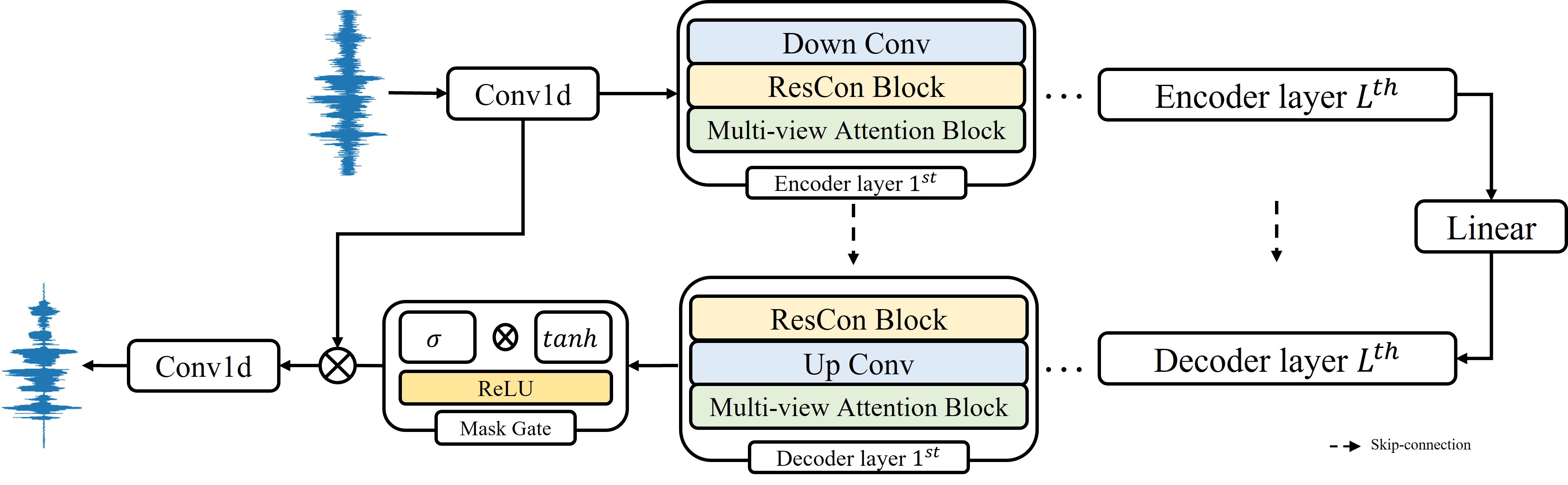}}
\caption{The overall architecture of MANNER.}
\label{fig:manner}
\end{figure*}

\section{MANNER}
\label{sec:method}

In this section, we introduce MANNER in detail. MANNER is based on an encoder-decoder consisting of a convolution layer, a convolution block, and an attention block.

\subsection{Encoder and Decoder}
\label{sec:encoder}

Before the encoder layer, we use a 1-D convolution layer, followed by batch normalization and ReLU activation, on the noisy input $x \in \mathbb{R}^{1 \times T}$, where $T$ is the signal length. The 1-D convolution layer expands the channel size according to $x \in \mathbb{R}^{N \times T}$, where $N$ denotes the channel size.
 
 As shown in Fig. \ref{fig:manner}, the encoder and decoder consist of $L$ layers containing Down and Up Conv layers, a Residual Conformer (ResCon) block, and a Multi-view Attention (MA) block. We use the linear transformation of the encoder output to pass the decoder layer. Each encoded output is connected with each decoding input by element-wise summation.

\noindent \textbf{Up \& Down Conv.} We use Down and Up Conv in the encoder and decoder, respectively. Down Conv, which reduces the signal length, consists of a convolution layer followed by batch normalization and ReLU activation. In contrast, Up Conv, which restores the signal to its original length, consists of a transposed convolution layer instead of a convolution layer. Up and Down Conv adjust the signal length with a kernel size of $K$ and a stride of $S$. We denote the signal length of each layer as $T_{l}$, where $l={1,2,...,L}$. 

\noindent \textbf{Mask Gate.} We obtain the mask $m \in \mathbb{R}^{N \times T}$ by applying a mask gate to the decoder output. The mask is estimated by the multiplication between the sigmoid and hyperbolic activation on the output, followed by ReLU activation. A convolution layer is always used before each activation function. We obtain the denoised $x' \in \mathbb{R}^{N \times T}$ through element-wise multiplication between the mask and the output of the first convolution layer, $x \in \mathbb{R}^{N \times T}$. Finally, the enhanced speech is obtained by applying the convolution layer, which reduces the channel size from $N$ to $1$, to the denoised $x'$.

\subsection{Residual Conformer block}
\label{sec:rescon}
\begin{figure}[h]
\centerline{\includegraphics[scale = 0.6]{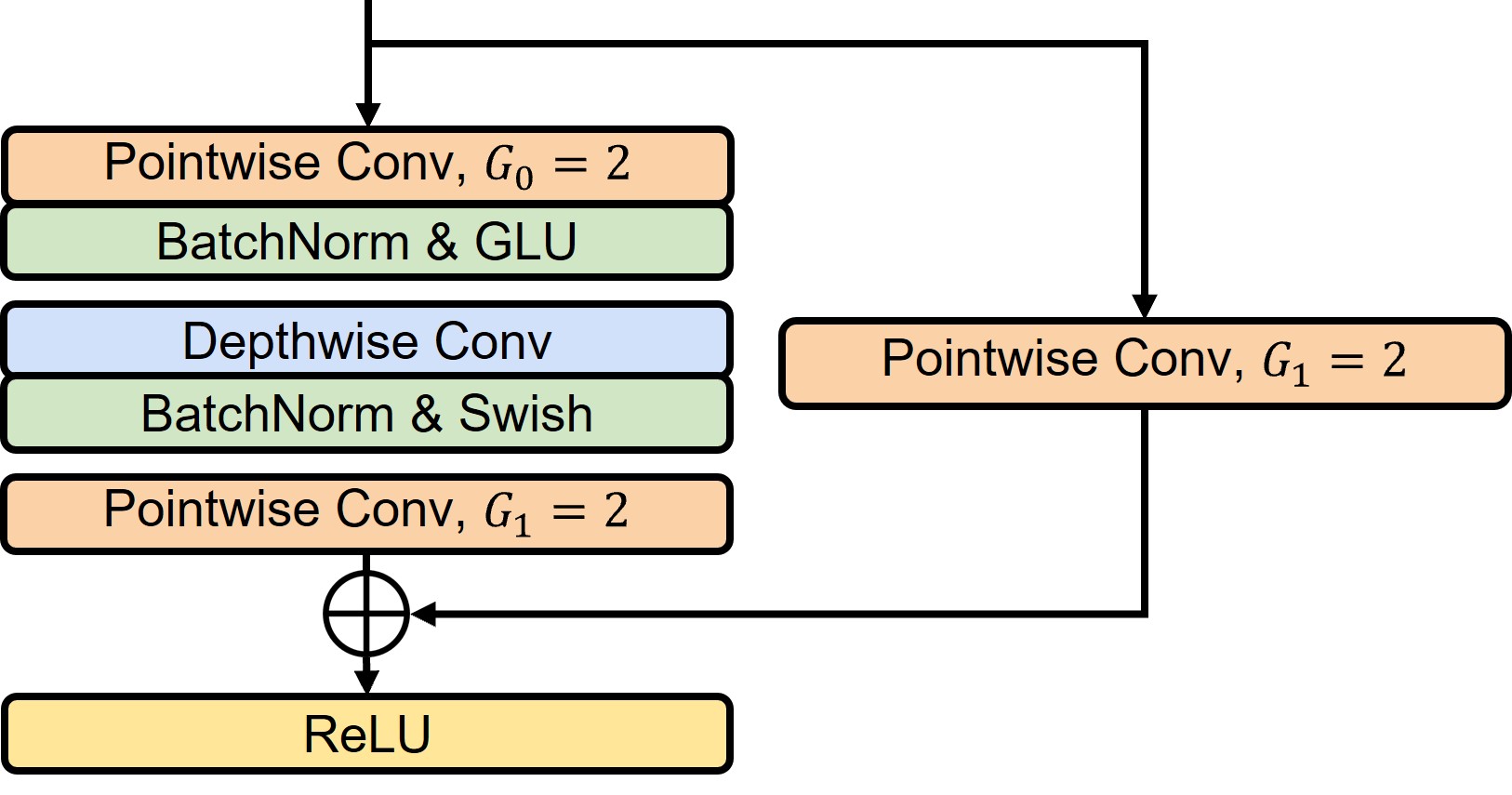}}
\caption{Residual Conformer block. $G_{0,1}$ indicates the channel growth rate of each pointwise convolution.}
\label{fig:rescon}
\end{figure}

Inspired by the efficient convolution block of Conformer \cite{gulati2020conformer}, we design a ResCon block to obtain enriched channel representation by expanding the channel size in deep layers. We modify the normalization and add a residual connection using a convolution layer. In addition, we redesign the method used to adjust the channel size. As shown in Fig. \ref{fig:rescon}, pointwise and depthwise convolution layers are followed by normalization and the activation function. $G_{1}$ adjusts the final channel size in the block, and we set $G_{1}=2$ and $G_{1}=1/2$ for the encoder and decoder layers, respectively. 
\begin{figure}[h]
\centerline{\includegraphics[scale = 0.5]{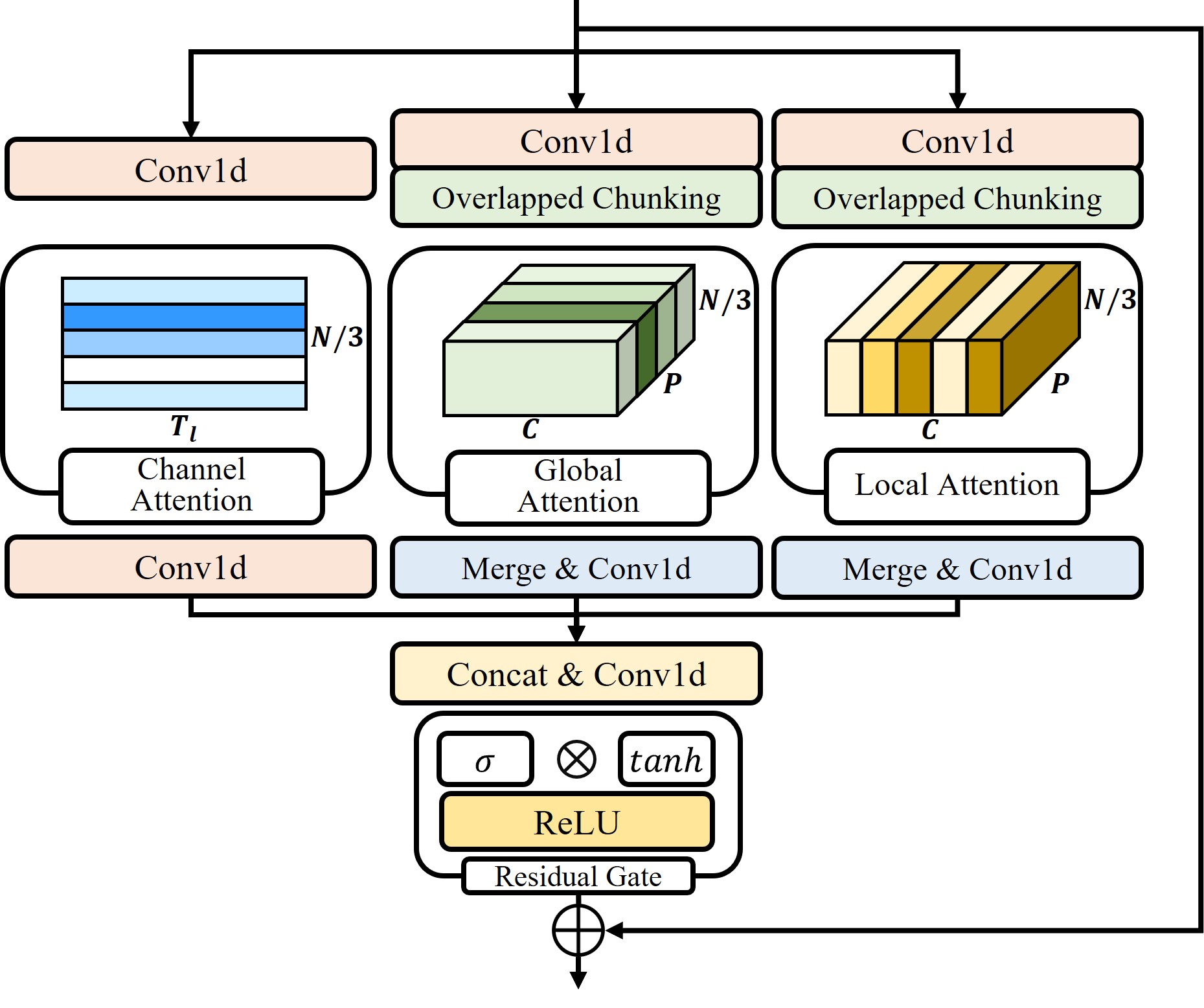}}
\caption{Multi-view Attention block.}
\label{fig:mablock}
\end{figure}
\subsection{Multi-view Attention block}
\label{sec:attention}
We design a MA block consisting of channel, global, and local attention to fully represent the signal information. Channel attention emphasizes representations from compressed channels. Global and local attention based on dual paths efficiently reflect long sequential features. In the MA block, the input passes through three paths consisting of a convolution layer that adjusts the channel size from $N$ to $N/3$. For global and local attention paths, we adopt chunking with an overlap ratio of 50\% to split $x \in \mathbb{R}^{N/3 \times T_{l}}$ into $x \in \mathbb{R}^{N/3 \times P \times C}$, where $P$ and $C$ denote the number of chunks and chunk size, respectively. By separating the global and local information, we can efficiently represent long sequential features. \\
\noindent \textbf{Channel Attention.} We adopt the channel attention used in \cite{woo2018cbam} to emphasize channel-wise representations. To aggregate the signal information, we apply average and max pooling to $x_{C} \in \mathbb{R}^{N/3 \times T_{l}}$, where $x_{C}$ is the input of the channel attention path after a convolution layer. Each pooling output passes through shared linear layers. The channel attention weight $\alpha_{C} \in \mathbb{R}^{N/3 \times 1}$ is estimated as follows:
\begin{equation}
\alpha_{C}= \sigma(W_{1}(W_{0}(x_{C}^{avg})) + W_{1}(W_{0}(x_{C}^{max})))
\label{eq:channel attention}
\end{equation}
\noindent where each weight is $W_{0}\in \mathbb{R}^{N/3 \times N/6}$ and $W_{1}\in \mathbb{R}^{N/6 \times N/3}$. The channel attention output is defined as $x_{C}'=x_{C} \times \alpha_{C}$.
\noindent \textbf{Global Attention.} We propose global attention based on the self-attention of Transformer \cite{vaswani2017attention}. To extract global sequential information, global attention considers chunk-wise representations in the chunked input $x_{G} \in \mathbb{R}^{N/3 \times P \times C}$. The global attention weight, $\alpha_{G} \in \mathbb{R}^{N/3 \times P \times P}$, and the output of global attention, $x_{G}' \in \mathbb{R}^{N/3 \times P \times C}$, are obtained based on self-attention, where $d_{k}$ is the chunk size for scaling.
\begin{equation}\label{eq:global attention}
  \begin{gathered}
    \alpha_{G}=softmax(\frac{QK^\top}{\sqrt{d_k}}) \\
    x_{G}' = W(\alpha_{G}V)
  \end{gathered}
\end{equation}
\noindent $Q, K$, and $V \in \mathbb{R}_{q,k,v}^{N/3 \times P \times C}$ are represented by linear transformation with each weight, $W_{q,k,v} \in \mathbb{R}^{1 \times C \times C}$, and $x_{G} \in \mathbb{R}^{N/3 \times P \times C}$. Finally, we apply a linear layer, $W\in \mathbb{R}^{1 \times C \times C}$, to $\alpha_{G}V$ to obtain the global attention output.

\noindent \textbf{Local Attention.} Local attention represents the local sequential features in each chunk. We design local attention using convolution layers to reduce the model complexity compared to self-attention. By adopting a small chunk size and large kernel size, the convolution layer can sufficiently represent local sequential features. We use a depthwise convolution layer with a kernel size of $C/2-1$ on the chunked input $x_L \in \mathbb{R}^{P \times N/3 \times C}$. After the depthwise convolution layer, we estimate the local attention weight $\alpha_{L} \in \mathbb{R}^{P \times 1 \times C} $ by concatenating the channel-wise average and max pooling as follows:
\begin{equation}
\alpha_{L}= \sigma(F([x_{L}^{avg};x_{L}^{max}]))
\label{eq:local attention}
\end{equation}
\noindent where $F$ is the convolution layer reducing the channel size from 2 to 1. Finally, we represent the output as $x_{L}'=x_{L} \times \alpha_{L}$. \\
\indent For global and local outputs, we merge the chunked data. After the three-path attention, we concatenate each output and pass it through a convolution layer. We apply the mask gate process as the residual gate to adjust the amount of information flow, followed by a residual connection.
\subsection{Loss function}
\label{sec:loss}
We combine L1 loss (time) and multi-resolution STFT loss (time-frequency) \cite{defossez2020real, yamamoto2020parallel} to optimize the model. We adopt the STFT loss of \cite{defossez2020real}, which is the sum of the spectral convergence and magnitude loss. To obtain the spectral convergence and magnitude loss, the $Frobenius$ and $L_{1}$ norms are applied, respectively. The loss of the clean and estimated speech is the sum of the $L_{1}$ and multi-resolution STFT loss as follows:
\begin{equation}\label{eq:loss}
  \begin{gathered}
    loss_{STFT}(y, \hat{y})= \frac{\lVert |STFT(y)| - |STFT(\hat{y})| \rVert_{F}}{\lVert |STFT(y)| \rVert_{F}} \\
    + \frac{1}{T}\lVert log(|STFT(y)|) - log(|STFT(\hat{y})|) \rVert_{1} \\
    loss(y, \hat{y})= \frac{1}{T}\lVert y-\hat{y} \rVert_{1}+\frac{1}{R}\sum_{r=1}^{R}loss_{STFT}^{r}(y,\hat{y})
  \end{gathered}
\end{equation}
\noindent where $y$ and $\hat{y}$ are the clean and estimated speech. The $loss_{STFT}^{r}$ indicates the STFT loss of different resolutions with combinations of hyperparameter (i.e., window lengths, hop sizes, FFT bins), as in \cite{defossez2020real}. \\
\indent We also apply the weighted loss \cite{choi2019phaseaware} to consider both clean and noise loss. Given that $n$ is the noise, the input signal is defined as $x=y+n$. The total loss of the proposed model is as follows, where $\hat{n}=x-\hat{y}$.
\begin{equation}
loss_{total}(x, y, \hat{y})=\alpha loss(y, \hat{y})+(1-\alpha)loss(n, \hat{n})
\label{eq:total loss}
\end{equation}
\noindent The weight $\alpha$ is defined as $\alpha=\lVert y \rVert^{2}_{2}/(\lVert y \rVert^{2}_{2}+\lVert n \rVert^{2}_{2})$, adjusting the ratio between the clean and noise speech.

\section{Experiments}
\label{sec:exp}

\subsection{Datasets}
\label{sec:data}

We evaluate MANNER on the VoiceBank-DEMAND dataset \cite{valentini2017noisy} by mixing the VoiceBank Corpus and DEMAND dataset. The train set consists of 11,572 utterances (14 male and 14 female) mixed with noise data with four signal-to-noise ratios (SNRs) (15, 10, 5, and 0 dB). The test set consists of 824 utterances (one male and one female) mixed with unseen noise data with four SNRs (17.5, 12.5, 7.5, and 2.5 dB). We use two speakers from the train set as the validation set. The data are downsampled from 48 kHz to 16 kHz for a fair comparison.
\begin{table*}
\caption{Comparison results on the VoiceBank-DEMAND dataset in terms of objective speech quality metrics.}\label{table:comparison}
\centering
\begin{tabular}{M{3cm}M{1.5cm}M{2cm}M{2cm}M{2cm}M{2cm}M{2cm}} 
\toprule
Model &  Domain  & PESQ & STOI(\%) & CSIG & CBAK & COVL \\
\midrule
SEGAN \cite{pascual2017segan} & T & 2.16 & - & 3.48 & 2.94 & 2.80 \\
Wave U-Net \cite{macartney2018improved} & T & 2.40 & - & 3.52 & 3.24 & 2.96 \\
PHASEN \cite{yin2020phasen} & T-F & 2.99 & - & 4.21 & 3.55 & 3.62 \\
SN-Net \cite{zheng2020interactive} & T-F & 3.12 & - & 4.39 & 3.60 & 3.77 \\
DEMUCS (large) \cite{defossez2020real} & T & 3.07 & 95 & 4.31 & 3.4 & 3.63 \\ 
MetricGAN+ \cite{fu2021metricgan+} & T-F & 3.15 & - & 4.14 & 3.16 & 3.64 \\
TSTNN \cite{wang2021tstnn} & T & 2.96 & 95 & 4.33 & 3.53 & 3.67 \\ 
\midrule
MANNER (small) & T & 3.12 & 95 & 4.45 & 3.61 & 3.82 \\
\textbf{MANNER} & T & \textbf{3.21} & \textbf{95} & \textbf{4.53} & \textbf{3.65} & \textbf{3.91} \\
\bottomrule
\end{tabular}
\end{table*}

\subsection{Evaluation metrics}
\label{sec:metric}
We adopt five objective measures to evaluate MANNER and the previous models. Perceptual Evaluation of Speech Quality (PESQ) \cite{recommendation2001perceptual} with a score ranging from -0.5 to 4.5 is used to evaluate speech quality. Short-time objective intelligibility (STOI) \cite{taal2011algorithm} with a score ranging from 0 to 100 is for speech intelligibility. We also consider three mean opinion score (MOS)-based measures whose scores ranging from 1 to 5 \cite{hu2007evaluation}. CSIG is the MOS prediction of the signal distortion, CBAK is the MOS prediction of the noise intrusiveness, and COVL is the MOS prediction of the overall signal quality.

\subsection{Implementation details}
\label{sec:implementation}
For training, we segment the signal into 4 seconds with a 1-second overlap and set a batch size of 4. We train MANNER for 300 epochs and maintain the best weights based on the validation score. Furthermore, we adopt the Adam optimizer and OneCycleLR scheduler to optimize the model. We set $lr_{min}=10^{-5}$ and $lr_{max}=10^{-2}$ for the OneCycleLR scheduler adjusting the learning rate during each epoch. During training, we vary the tempo of the signal within the range of 90\% to 110\% \cite{ko2015audio}. For MANNER, we use $K=8, S=4, N=60, L=4$, and $C=64$. To verify the performance and efficiency, we also include MANNER (small) in the comparison, containing MA block only in the $L^{th}$ layer and using the same parameters as MANNER.
\begin{figure}[h]
\centerline{\includegraphics[scale = 0.35]{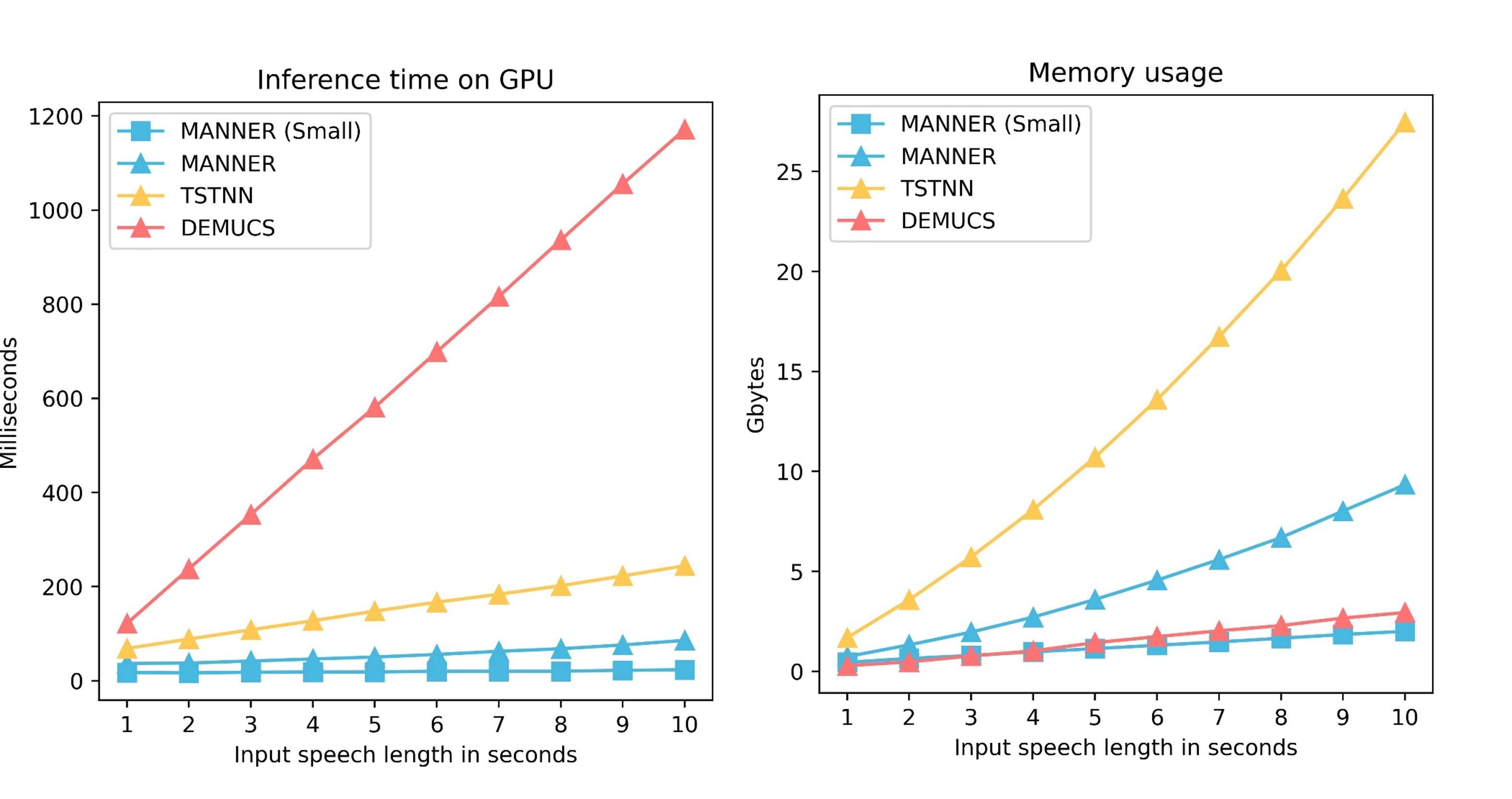}}
\caption{Efficiency comparison performed on the same machine with RTX A6000 GPU.}
\label{fig:efficiency comparison}
\end{figure}
\subsection{Experimental results}
\label{sec:experiment results}

We compared the proposed models with existing models, including time and time-frequency domain methods. As shown in Table \ref{table:comparison}, MANNER achieves state-of-the-art performance in terms of five objective speech quality measures. Although MANNER (small) does not achieve the best performance, it still outperforms the previous methods.

To verify the efficiency of the proposed models, we compared them with the time domain methods, DEMUCS \cite{defossez2020real} and TSTNN \cite{wang2021tstnn}, in terms of inference speed and memory usage. We measured these quantities with the signal length set from 1 to 10 seconds. Fig. \ref{fig:efficiency comparison} shows that MANNER has high inference speed and relatively low memory usage compared to the previous methods. In addition, MANNER (small) achieves not only higher performance than the previous methods, but also the highest efficiency.
\begin{table}[h]
\centering
\caption{Comparison results depending on attention types and weighted loss (wLoss).}
\label{table:ablation}
\resizebox{0.473\textwidth}{!}{%
\begin{tabular}{c|cccc|c}
\toprule
Ver. & wLoss. & Channel Att. & Global Att. & Local Att. & PESQ\\
\midrule
Base& & & & & 3.00 \\
1& \checkmark & & & & 3.04\\
2& \checkmark & \checkmark & & \checkmark & 3.12\\
3& \checkmark & \checkmark & \checkmark & & 3.16\\
4& \checkmark & & \checkmark & \checkmark & 3.18\\
\midrule
MANNER& \checkmark & \checkmark & \checkmark & \checkmark & \bf 3.21\\
\bottomrule
\end{tabular}}
\end{table}
\subsection{The influence of attention block and loss}
\label{sec:ablation}

We conducted an ablation experiment to understand the influence of the proposed attention block and weighted loss on MANNER’s performance. We examined the effects of each component of the proposed methods. Table \ref{table:ablation} shows that each attention and weighted loss contributes to the improvement of performance. The result of Ver. 4 suggests the importance of considering long signal information, but considering all views of the signal is necessary to achieve higher performance. 

\section{Conclusion}
\label{sec:con}
In this study, we proposed MANNER, which efficiently represents channel and long sequential features of the signal, designed for speech enhancement in the time domain. MANNER’s results on the VoiceBank-DEMAND dataset highlight that MANNER achieves state-of-the-art performance compared to existing models. In addition, MANNER (small) is superior to previous time-domain methods in terms of performance and efficiency. Finally, the ablation experiment suggests that it is important to consider all representations of the signal and optimize both clean and noise loss.

\bibliographystyle{IEEEbib}
\bibliography{paper}

\begin{thebibliography}{10}

\bibitem{choi2019phaseaware}
H.-S.Choi et~al.,
\newblock ``Phase-aware speech enhancement with deep complex u-net,'' 2019.

\bibitem{fu2019metricgan}
S.-W.Fu et~al.,
\newblock ``Metricgan: Generative adversarial networks based black-box metric
  scores optimization for speech enhancement,''
\newblock in {\em ICML}, 2019.

\bibitem{fu2021metricgan+}
S.-W.Fu et~al.,
\newblock ``Metricgan+: An improved version of metricgan for speech
  enhancement,''
\newblock {\em arXiv preprint arXiv:2104.03538}, 2021.

\bibitem{yin2020phasen}
D.Yin, C.Luo, Z.Xiong, and W.Zeng,
\newblock ``Phasen: A phase-and-harmonics-aware speech enhancement network,''
\newblock in {\em Proc. AAAI}, 2020, vol.~34, pp. 9458--9465.

\bibitem{zheng2020interactive}
C.Zheng, X.Peng, Y.Zhang, S.Srinivasan, and Y.Lu,
\newblock ``Interactive speech and noise modeling for speech enhancement,''
\newblock {\em arXiv preprint arXiv:2012.09408}, 2020.

\bibitem{wang2021tstnn}
K.Wang, B.He, and W.-P.Zhu,
\newblock ``Tstnn: Two-stage transformer based neural network for speech
  enhancement in the time domain,''
\newblock in {\em ICASSP}. IEEE, 2021, pp. 7098--7102.

\bibitem{rethage2018wavenet}
D.Rethage, J.Pons, and X.Serra,
\newblock ``A wavenet for speech denoising,''
\newblock in {\em ICASSP}. IEEE, 2018, pp. 5069--5073.

\bibitem{pandey2020dual}
A.Pandey and D.Wang,
\newblock ``Dual-path self-attention rnn for real-time speech enhancement,''
\newblock {\em arXiv preprint arXiv:2010.12713}, 2020.

\bibitem{defossez2020real}
A.Defossez, G.Synnaeve, and Y.Adi,
\newblock ``Real time speech enhancement in the waveform domain,''
\newblock {\em arXiv preprint arXiv:2006.12847}, 2020.

\bibitem{macartney2018improved}
C.Macartney and T.Weyde,
\newblock ``Improved speech enhancement with the wave-u-net,''
\newblock {\em arXiv preprint arXiv:1811.11307}, 2018.

\bibitem{pascual2017segan}
S.Pascual, A.Bonafonte, and J.Serra,
\newblock ``Segan: Speech enhancement generative adversarial network,''
\newblock {\em arXiv preprint arXiv:1703.09452}, 2017.

\bibitem{pandey2020densely}
A.Pandey and D.Wang,
\newblock ``Densely connected neural network with dilated convolutions for
  real-time speech enhancement in the time domain,''
\newblock in {\em ICASSP}. IEEE, 2020, pp. 6629--6633.

\bibitem{pandey2021dense}
A.Pandey and D.Wang,
\newblock ``Dense cnn with self-attention for time-domain speech enhancement,''
\newblock {\em IEEE/ACM Transactions on Audio, Speech, and Language
  Processing}, vol. 29, pp. 1270--1279, 2021.

\bibitem{hsieh2020wavecrn}
T.-A.Hsieh, H.-M.Wang, X.Lu, and Y.Tsao,
\newblock ``Wavecrn: An efficient convolutional recurrent neural network for
  end-to-end speech enhancement,''
\newblock {\em IEEE Signal Processing Letters}, vol. 27, pp. 2149--2153, 2020.

\bibitem{ronneberger2015u}
O.Ronneberger, P.Fischer, and T.Brox,
\newblock ``U-net: Convolutional networks for biomedical image segmentation,''
\newblock in {\em ICM}. Springer, 2015, pp. 234--241.

\bibitem{luo2020dual}
Y.Luo, Z.Chen, and T.Yoshioka,
\newblock ``Dual-path rnn: efficient long sequence modeling for time-domain
  single-channel speech separation,''
\newblock in {\em ICASSP}. IEEE, 2020, pp. 46--50.

\bibitem{chen2020dual}
J.Chen, Q.Mao, and D.Liu,
\newblock ``Dual-path transformer network: Direct context-aware modeling for
  end-to-end monaural speech separation,''
\newblock {\em arXiv preprint arXiv:2007.13975}, 2020.

\bibitem{subakan2021attention}
C.Subakan, M.Ravanelli, S.Cornell, M.Bronzi, and J.Zhong,
\newblock ``Attention is all you need in speech separation,''
\newblock in {\em ICASSP}. IEEE, 2021, pp. 21--25.

\bibitem{gulati2020conformer}
A.Gulati et~al.,
\newblock ``Conformer: Convolution-augmented transformer for speech
  recognition,''
\newblock {\em arXiv preprint arXiv:2005.08100}, 2020.

\bibitem{woo2018cbam}
S.Woo, J.Park, J.-Y.Lee, and I.~S.Kweon,
\newblock ``Cbam: Convolutional block attention module,''
\newblock in {\em Proc. ECCV}, 2018, pp. 3--19.

\bibitem{vaswani2017attention}
A.Vaswani, N.Shazeer, N.Parmar, J.Uszkoreit, L.Jones, A.~N.Gomez, {\L}.Kaiser,
  and I.Polosukhin,
\newblock ``Attention is all you need,''
\newblock in {\em Advances in neural information processing systems}, 2017, pp.
  5998--6008.

\bibitem{yamamoto2020parallel}
R.Yamamoto, E.Song, and J.-M.Kim,
\newblock ``Parallel wavegan: A fast waveform generation model based on
  generative adversarial networks with multi-resolution spectrogram,''
\newblock in {\em ICASSP}. IEEE, 2020, pp. 6199--6203.

\bibitem{valentini2017noisy}
C.Valentini-Botinhao et~al.,
\newblock ``Noisy speech database for training speech enhancement algorithms
  and tts models,''
\newblock 2017.

\bibitem{recommendation2001perceptual}
I.-T.Recommendation,
\newblock ``Perceptual evaluation of speech quality (pesq): An objective method
  for end-to-end speech quality assessment of narrow-band telephone networks
  and speech codecs,''
\newblock {\em Rec. ITU-T P. 862}, 2001.

\bibitem{taal2011algorithm}
C.~H.Taal, R.~C.Hendriks, R.Heusdens, and J.Jensen,
\newblock ``An algorithm for intelligibility prediction of time--frequency
  weighted noisy speech,''
\newblock {\em IEEE Transactions on Audio, Speech, and Language Processing},
  vol. 19, no. 7, pp. 2125--2136, 2011.

\bibitem{hu2007evaluation}
Y.Hu and P.~C.Loizou,
\newblock ``Evaluation of objective quality measures for speech enhancement,''
\newblock {\em IEEE Transactions on audio, speech, and language processing},
  vol. 16, no. 1, pp. 229--238, 2007.

\bibitem{ko2015audio}
T.Ko, V.Peddinti, D.Povey, and S.Khudanpur,
\newblock ``Audio augmentation for speech recognition,''
\newblock in {\em Sixteenth annual conference of the international speech
  communication association}, 2015.

\end{thebibliography}

\end{document}